\documentclass[11pt,paper]{article}
\usepackage{jheppub}

\usepackage{float}
\usepackage[usenames,dvipsnames,table]{xcolor}
\usepackage{graphicx,amsmath,amssymb,multirow,array,bm,mathrsfs}
\usepackage{epsf,amsfonts,slashed}
\usepackage[numbers,sort&compress]{natbib}

\title{
An anomalous propulsion mechanism using magnetic fields
}
\author{Evgeny Shaverin}

\affiliation{Department of Physics, Technion, Haifa 32000, Israel}

\emailAdd{evgeny@campus.technion.ac.il}
\abstract{
We consider a gas composed of a single family of standard model leptons which are approximately massless and trapped inside a charged rotating shell. Due to the magnetic vortical effect, the leptons gain momentum in the direction of the magnetic field induced by the rotating shell. We compute this momentum gain in a perturbative expansion and discuss the possible application of it to pulsar kicks.
}
\begin{document}
\maketitle

\section{Introduction }
\label{S:intro}

Quantum mechanical effects are usually manifest at very small scales. Yet, there is mounting evidence that flavor and gravitational anomalies can have significant repercussions on hydrodynamic behavior of relativistic fluids \cite{Erdmenger:2008rm,Banerjee:2008th,Son:2009tf,Neiman:2010zi,Landsteiner:2011cp,Landsteiner:2012kd,Banerjee:2012iz,Jensen:2012kj,Haehl:2013hoa,Jensen:2013kka,Jensen:2013rga,Golkar:2015oxw,Chowdhury:2016cmh,Glorioso:2017lcn}. Such hydrodynamic phenomenon may be responsible for charge asymmetry of hadrons emitted as a result of a collision of two heavy ions \cite{Kharzeev:2007tn,Kharzeev:2007jp,Kharzeev:2013ffa,Yin:2015fca,Kharzeev:2015znc} and there are candidate setups for measuring these phenomenon in condensed matter systems \cite{Loganayagam:2012pz,Basar:2013iaa,Chernodub:2013kya,Landsteiner:2013sja,Yamamoto:2015fxa,Zhang:2015gwa,Sen:2016jzl,Lucas:2016omy} and astrophysics \cite{Charbonneau:2009ax,Charbonneau:2010jx,Kaminski:2014jda,Shaverin:2014xya,Yamamoto:2015gzz,Yamamoto:2016zut}. 

One of the central observations which have paved the road towards a complete understanding of the role of anomalies in hydrodynamic behavior was carried out in \cite{Son:2009tf}. Following the observations of \cite{Erdmenger:2008rm,Banerjee:2008th} the authors of \cite{Son:2009tf} have demonstrated that currents associated with symmetries which possess 't Hooft anomalies in $3+1$ dimensions have special characteristics when in or near thermodynamic equilibrium. For instance, when considering linear response, such currents react to vorticity of the fluid they are in, or to an external flavor magnetic field. These phenomenon are often referred to as the chiral vortical effect or the chiral magnetic effect respectively. Generalizations to gravitational and mixed anomalies, to triangle (ABJ) anomalies, and to other dimensions can be found throughout the literature \cite{Loganayagam:2011mu,Kharzeev:2011ds,Jensen:2012kj,Banerjee:2013fqa,Jensen:2013rga,Jensen:2013kka}.

The relation between anomalies and hydrodynamic behavior of currents which generate anomalous symmetries raises the possibility of a macroscopic manifestation of anomalies. Unfortunately, to actually observe the anomalous behavior of currents one needs good control over a macroscopic system whose dynamics are well approximated by relativistic hydrodynamics and whose underlying fundamental fields include chiral fermions. One option for generating such configurations is in condensed matter systems, Weyl semi-metals in particular, whose low energy effective description includes chiral fermions. 

In the context of particle physics, lepton number is anomalous (assuming a single, massless, left handed neutrino per family) and so, raises the possibility of observing the effects predicted by \cite{Son:2009tf} in an appropriate astrophysical setup. Indeed in a theory comprised of a single family of leptons with approximately massless electrons, the left handed lepton number current and the right handed lepton number current will be separately conserved from a classical standpoint but are anomalous due to quantum effects. As argued for in \cite{Kaminski:2014jda}, if such a gas is placed inside a magnetized vessel then, upon its release, the chiral magnetic effect will force the gas to flow in the direction of the magnetic field and, due to momentum conservation, provide the vessel with linear momentum in a direction opposite that of the released gas. A similar propulsion mechanism associated with the chiral vortical effect was discussed in \cite{Shaverin:2014xya}.

In \cite{Kaminski:2014jda} it was suggested that the magnetically driven propulsion mechanisms may be used to explain pulsar kicks (see also the discussion in \cite{Shaverin:2014xya}). Pulsars are rotating and highly magnetized neutron stars generated when a star collapses under its own pressure and releases a burst of neutrinos. Young pulsars are observed to travel at very high velocities ranging up to a thousand kilometers per second relative to the progenitor star \cite{Cordes:1997mm,Chatterjee:2005mj}. The explanation of these irregular velocities is currently not agreed on by the scientific community (see, e.g., \cite{Lai:1999za}) and the above phenomenon is referred to as ``pulsar kicks''.

In this work we offer a controlled setting where the magnetically driven anomalous propulsion mechanism of \cite{Kaminski:2014jda} may be studied. We consider a hydrostatically equilibrated gas comprised of a single family of leptons with massless neutrinos at a temperature much larger than the mass of the electrons, trapped inside a spherical rotating charged shell. Neglecting the response of the shell to the gas we may use the techniques developed in \cite{Banerjee:2012iz,Jensen:2012kj,Jensen:2013kka,Jensen:2013rga} to compute (perturbatively) the response of the stress-tensor of the gas to the magnetic field generated by the rotating charged shell. Unsurprisingly, we find that due to the chiral magnetic effect, the stress tensor of the equilibrated gas is inhomogenous with stronger support in the direction of the magnetic field. If the shell becomes instantaneously transparent then momentum conservation implies that the gas and shell will be thrust in opposite directions generating an anomaly based propulsion mechanism. Within our setup we estimate the dependence of final momentum of the shell on the radius, mass, charge and angular velocity (or magnetic field) of the shell.


Needless to say, our simplistic model, described above, is far removed from any sort of realistic description of neutron star formation. Nevertheless, given the analytic control we have over it, it seems a waste not to put it to use, at least as an initial estimate for the viability of the anomalous propulsion mechanism to generate pulsar kicks. 
Our naive analysis implies that the magnetic based propulsion mechanism is a possible candidate for pulsar kicks only if the magnetic fields during the collapse process are extremely high or that during the collapse process the radius of the core is close enough to the Schwarzschild radius.

Let us comment that the propulsion mechanism we propose is similar but not quite the same as a jet propulsion mechanism described in \cite{Kaminski:2014jda}. In a jet propulsion mechanism the jet engine continuously generates gas which propels the vessel forward. In our model there is a limited amount of gas which is released instantaneously, the effect being somewhat similar to propelling a balloon by punching a hole through it (though the directionality in our case is a result of the anomaly and not of directional release of the gas). 
It seems to us more likely that a balloon type mechanism rather than a jet propulsion mechanism is a good model for describing pulsar kicks. In core collapse supernova, prior to neutrino emission the neutrinos are trapped inside the star due to their short mean free path during which they presumably thermalize with the rest of the trapped matter. In the course of the collapse process there is a period of roughly one second in which the neutrinos are released due to an increase in their mean free path. But once the mean free path of the neutrinos is larger than the radius of the core it seems difficult to conceive of a mechanism by which thermal neutrinos will be continuously generated during this time. 

\section{Hydrostatic equilibrium inside a charged spinning shell}

Recall that the dynamics of a gas of relativistic particles is characterized by a handful of hydrodynamic fields: a temperature field $T$, a velocity field $u^{\mu}$ and chemical potentials $\mu_i$ conjugate to the conserved charges. The dynamical equations for these fields are given by energy momentum conservation and charge conservation. For example, in the absence of charge the dependence of the stress tensor on the hydrodynamic variables is given by
\begin{equation}
	T^{\mu\nu} = \epsilon(T ) u^{\mu}u^{\nu} + P(T ) (g^{\mu\nu} + u^{\mu}u^{\nu}) + \mathcal{O}(\partial)\,,
\end{equation}
where $\mathcal{O}(\partial)$ includes expressions which contain derivatives of the hydrodynamic fields, $\epsilon$ is the energy density and $P$ the pressure. Energy-momentum conservation will then provide us with equations of motion for $u^{\mu}$ (assumed to be unit normalized) and $T$.

In what follows we wish to consider a gas of a single family of standard model particles with massless leptons, trapped inside a stationary spherical rotating charged shell. Thus, the classically conserved charges which will be relevant for this work include the energy momentum tensor $T^{\mu\nu}$ and three $U(1)$ charge currents: the electric charge current $J_e^{\mu}$, and the left and right lepton number charge $J_r^{\mu}$ and $J_{\ell}^{\mu}$ which are separately conserved in a classical theory of massless electrons. As pointed out in \cite{Kaminski:2014jda}, while the three Abelian currents mentioned above are classically conserved, the left and right lepton number currents are anomalous. We rederive this result in appendix \ref{A:triangle}.

If we are to take into account the gravitational and electric field induced by the motion of the gas then in addition to the conservation laws for the stress tensor and electric current we should also include the dynamics of the metric and Maxwell field. Such a system of equations is often referred to as gravito- or magneto-hydrodynamics and has been recently reconsidered in \cite{Grozdanov:2016tdf,Kovtun:2016lfw,Hernandez:2017mch}. In this work we will assume that the gas interacts weakly with a background electromagnetic field and metric so that the dynamics of the metric and gauge field decouple from that of the gas. 

Since the background metric and gauge field are stationary, the gas will be in hydrostatic equilibrium. Indeed, when a gas is placed in a time independent background metric $g_{\mu\nu}$ and external electromagnetic potential $A_{\mu}$ it will reach an equilibrium state which we refer to as hydrostatic equilibrium. As argued in \cite{Jensen:2012jh,Banerjee:2012iz}, in hydrostatic equilibrium, the constitutive relations are such that
\begin{equation} \label{E:Solution}
	T=\frac{T_{0}}{\sqrt{-g_{tt}}}\,,\quad 
	u^{\mu}=\frac{\delta_{t}^{\mu}}{\sqrt{-g_{tt}}}\,,\quad
	\mu_i = \frac{T}{T_{0}}A_{t\,i} \,,
\end{equation}
solve the conservation equations at any order in the derivative expansion. Here $T_0$ is the inverse parametric length of the Euclidean time circle and $A_i$ is an external source conjugate to the charges associated with the chemical potential $\mu_i$. 

Following \cite{Jensen:2012kj,Jensen:2013kka,Jensen:2013rga} the constitutive relations for the stress tensor in hydrostatic equilibrium, expanded to first order in derivatives, are completely characterized by triangle anomalies (in the absence of anomalies the leading constitutive relations are second order in derivatives). They are given by
\begin{equation}
\label{E:Tmnanomaly}
	T^{\mu\nu} = \epsilon(T) u^{\mu}u^{\nu} + P(T) (g^{\mu\nu} + u^{\mu}u^{\nu}) + u^{\mu}q^{\nu} + u^{\nu}q^{\mu} + \mathcal{O}(\partial^2)\,,
\end{equation}
where 
\begin{align} 
\label{E:q_first_order}
	8\pi^2 q^{\mu}= & 2\left( q_r^3 \mu_{r}^{3} + q_e q_r^2 \mu \mu_{r}^{2} + q_e^2 q_r \mu^{2}\mu_{r} - 2 q_l^3 \mu_{\ell}^{3} - q_e q_l^2 \mu\mu_{\ell}^{2} - q_e^2 q_l \mu^{2}\mu_{\ell} + \frac{8\pi^{2}q_r }{24}\mu_{r}T^{2} - \frac{8\pi^{2}q_l}{12} \mu_{\ell}T^{2}\right)\omega^{\mu} \nonumber \\
	& + \left( 3 q_r^3 \mu_{r}^{2} + 2 q_e q_r^2 \mu\mu_{r} + q_e^2 q_r \mu^{2} + \frac{8\pi^{2}q_r}{24} T^{2}\right)B_{r}^{\mu}
	-\left(6 q_l^3\mu_{\ell}^{2}+2q_eq_l^2 \mu\mu_{\ell}+q_e^2 q_l \mu^{2} + \frac{ 8\pi^{2} q_l}{12} T^{2}\right)B_{\ell}^{\mu} \nonumber \\
	& +\left( q_eq_r^2 \mu_{r}^{2} +  2 q_e^2 q_r \mu\mu_{r} - q_e q_l^2 \mu_{\ell}^{2} - 2q_e^2 q_l \mu\mu_{\ell}\right)B^{\mu} \,,
\end{align}
with $B_i^{\mu}$ the magnetic field associated with the source $A_i$ and $\omega^{\mu}$ the vorticity,
\begin{equation}
\label{E:B_and_omega}
	B_i^{\mu}=\frac{1}{2}\epsilon^{\mu\nu\alpha\beta}u_{\nu}F_{i\alpha\beta}
	\qquad
	\omega^{\mu}=\epsilon^{\mu\nu\alpha\beta}u_{\nu}\nabla_{\alpha}u_{\beta} \,,
\end{equation}
with $F_{i\alpha\beta} = \partial_{\alpha} A_{i\,\beta} - \partial_{\beta} A_{i\,\alpha}$. To obtain \eqref{E:q_first_order} we have used the general algorithm described in \cite{Jensen:2012kj,Jensen:2013kka,Jensen:2013rga} which provides a method for evaluating constitutive relations which are uniquely determined by anomalies, directly from the relevant anomaly polynomial. A somewhat technical but detailed derivation of \eqref{E:q_first_order} can be found in appendix \ref{A:polynomial}.

Given \eqref{E:Tmnanomaly} and \eqref{E:Solution}, all that remains in order to compute the anisotropy of the stress tensor for our system is to evaluate the background metric and gauge field. The metric and electric field associated with a spherically rotating charged shell with angular velocity $\omega$ can be obtained by solving the Einstein Maxwell equations perturbatively in $\omega$, see \cite{Pfister}. In Appendix \ref{A:rotating} we have rederived this solution in a form convenient to this work. If we denote the parametric radius of the shell by $R$ its mass by $M$ and its charge by $Q$, we find that the line element inside the shell is given by
\begin{subequations}
\label{E:background}
\begin{equation} \label{IntMetric}
	ds^2 = -\frac{\left(R^{2}-r_{s}^{2}\right)^{2}}{\left(2r_{a}R+R^{2}+r_{s}^{2}\right)^{2}}dt^{2}+\frac{\left(2r_{a}R+R^{2}+r_{s}^{2}\right)^{2}}{R^{4}}\left[dr^{2}+r^{2}d\theta^{2}+r^{2}\sin^{2}\theta\left(d\phi-C_3 \omega dt\right)^{2}\right]
\end{equation}
where
\begin{equation} \label{E:ra&rs}
	r_a = M/2 \,,
	\qquad
	r_s = \frac{1}{2} \sqrt{M^2 - Q^2} \,,
\end{equation}
and $ C_3 $ is a spacetime independent constant which has been computed in Appendix \ref{A:rotating} (see equation \eqref{E:C1&C3}). The electromagnetic gauge potential is given by
\begin{equation} \label{E:A_mu}
	A_{\mu} = -\mu\delta_{\mu}^{t}-\frac{1}{2}\omega r^{2}C_{1}\sin^{2}\theta\delta_{\mu}^{\phi} 
\end{equation}
inside the shell, where $\mu$ is an integration constant and $C_1$ is a spacetime independent constant given in appendix \ref{A:rotating} (see equation \eqref{E:C1&C3}). 
Note that the magnetic field inside the shell is uniform and given by
\begin{equation}
\label{E:Bval}
	B = \omega C_1\,,
\end{equation}
in the longitudinal direction.
Since there are no electromagnetic left and right fermion field strengths we will use 
\begin{equation}
\label{E:ArandAl}
	A_{r\,\mu} = \mu_r \delta_{\mu}^t
	\qquad
	A_{\ell\,\mu} = \mu_{\ell} \delta_{\mu}^t
\end{equation}
\end{subequations}
for the external flavor fields associated with the right and left handed fermions. 

\begin{figure}[hbt]
\begin{center}
	\includegraphics[scale=0.85]{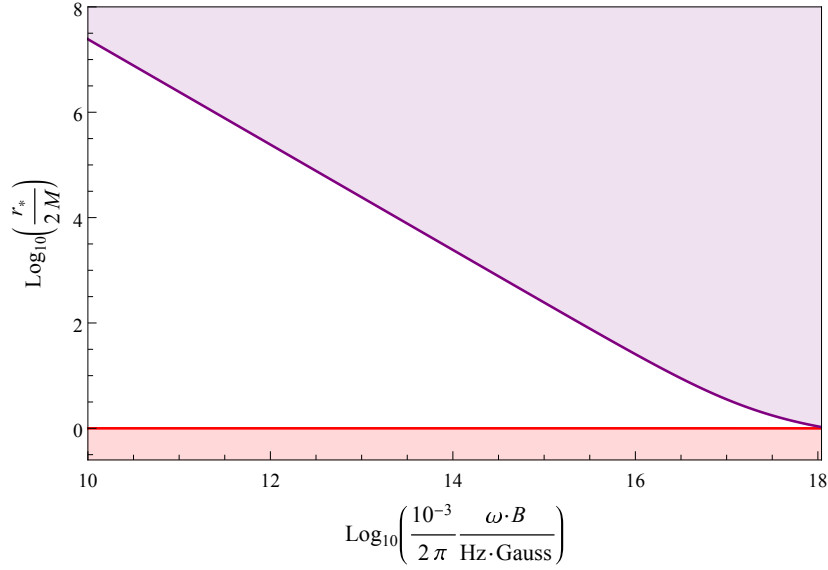}
	\protect\caption{\label{F:physical} The white region corresponds to the allowed range of $B$, $\omega$, $r_*$ and $M$ for the line element \eqref{E:background}. The solid red line represents $R<r_s$, below which black hole solutions are expected. The purple line represents solutions where $Q^2=M^2$ above which the solutions are hyper charged and the line element becomes complex. The vertical axis is measured in natural units where $G_N=c=1$.}
\end{center}
\end{figure}

As long as $R>r_s$ the interior of the shell is that of flat space. Once $R<r_s$ we expect a black hole to form and render our solution invalid. Likewise, if the charge of the shell is larger than its mass, $Q^2>M^2$, then the line element becomes complex and the solution is unphysical. We will refer to such configurations as hyper-charged. Thus, we will restrict ourselves to $R<r_s$ and $Q^2 < M^2$. It is convenient to trade $Q$ for the magnetic field in the interior $B$ and the parametric radius of the shell $R$ with its physical radius $r_*$, 
\begin{equation}
\label{Rstar}
	r_* = \int_0^R dr \sqrt{g_{rr}}\,.
\end{equation}
In figure \ref{F:physical} we have plotted the allowed values of $M$, $B$ and $r_*$ which follow from $R<r_s$ and $Q^2 < M^2$.

With the interior metric of the shell at hand we may evaluate the inhomogenous contributions to the stress tensor (at leading order in the derivative expansion) of the gas of leptons trapped inside it. Inserting \eqref{E:background} into \eqref{E:q_first_order} and using \eqref{E:Solution} we find
\begin{align} 
\begin{split}
\label{E:q_lep_m_e1}
		q^{z}= & -4cq\mu_{-}\left(\mu+\mu_{lep}\right)B-\frac{2}{3}c\Bigl[\mu_{lep}^{3}+3\mu_{-}\left(2\mu^{2}+\mu_{-}^{2}\right)  \\ 
		& +\,3\mu_{-}\mu_{lep}\left(4\mu+3\mu_{lep}+\mu_{-}\right)+\left(\mu_{lep}+3\mu_{-}\right)\pi^{2}T^{2}\Bigr]\Omega \\
		u^{t} = & \frac{2r_{a}R+R^{2}+r_{s}^{2}}{R^{2}-r_{s}^{2}} \,,  \\
\end{split}
\end{align}
where we have defined
\begin{equation}
\label{E:mulep&mum}
\mu_{lep}=\frac{\mu_{l}+\mu_{r}}{2}\,,\qquad\mu_{-}=\frac{\mu_{l}-\mu_{r}}{2}\,,
\end{equation}
the magnetic field $B$ is given by \eqref{E:Bval} and the vorticity by
\begin{equation}
	\Omega = \epsilon^{z\mu\nu\rho} u_{\mu} \partial_{\nu}u_{\rho}
\end{equation}
whose explicit form we will not need. The other components of $u^{\mu}$ and $q^{\mu}$ vanish.
Inserting \eqref{E:q_lep_m_e1} into \eqref{E:Tmnanomaly} we may compute the inhomogenous contributions to the momentum density $T^{0z}$.
We remind the reader that our result is valid for a gas of fermions (with massless neutrinos and with electrons whose mass is suppressed relative to the temperature) in an approximation where the shell does not backreact on the fermions. Working perturbatively in the electric charge or metric one may compute corrections to \eqref{E:q_lep_m_e1} (see, e.g., \cite{Jensen:2013vta}).

With the stress tensor for the trapped gas at hand we may compute its total momentum by integrating the momentum density,
\begin{equation} \label{PU}
	P^{\mu}=\int T^{\mu0}\sqrt{-g}\,d^{3}x\,.
\end{equation}
From \eqref{E:q_lep_m_e1} we find that
\begin{equation}
	P^z = P_{B}^z + P_{\Omega}^z\,,
\end{equation}
where, $P_B^z$ represents the contribution to the momentum coming from the magnetic field,
\begin{align} 
\begin{split}
\label{PBz}
	P_{B}^{z}=&2\pi\int\frac{4cq\mu_{-}\left(\mu+\mu_{lep}\right)\left(E_{\rho}g_{t\phi}-\omega\rho B_{z}g_{tt}\right)}{\left(-g_{tt}\right)\sqrt{g_{t\phi}^{2}-g_{tt}g_{\phi\phi}}}\sqrt{g_{\phi\phi}}d\rho dz \,,\\
	&=-16\pi c\,q\mu_{-}\left(\mu+\mu_{lep}\right)\omega C_{1}\frac{R^{3}\left(R^{2}+2r_{a}R+r_{s}^{2}\right)}{3\left(R^{2}-r_{s}^{2}\right)} \,,
\end{split}
\end{align}
and $P^z_{\Omega}$ represents the contribution to the momentum coming from the vorticity. As discussed in \cite{Shaverin:2014xya} the contribution of the vorticity is negligible unless an ergosphere is formed. In what follows we will focus on the contribution coming from the magnetic field. The other spatial components of the momentum vanish.

\section{A balloon-like propulsion mechanism}

The end result \eqref{PBz} specifies the momentum of a gas of a single family of standard model fermions trapped inside a charged rotating shell in an approximation where the gas does not interact with the shell and the electrons and a left handed neutrino are massless. In practice, once the gas interacts with the shell, the center of mass of the entire system must remain in equilibrium as it was before the gas thermalized implying a somewhat complex equilibrium configuration of the joint shell and gas system. Alternately, an external agent must hold the shell in place in order to keep the system equilibrated in our hydrostatic configuration. These unfortunate circumstances imply that it is difficult to estimate whether, upon release of the gas, the shell will gain momentum so as to travel at sufficiently high velocities, inline with current observations of pulsar kicks; in order for the shell to gain momentum due to losing the gas it contains there must exist an interaction term between the shell and gas. Nevertheless, with a lack of a full solution to the gravito-hydrodynamic equations of motion, we will treat the negative of \eqref{PBz} as a crude estimate of the momentum of the shell once the gas of leptons is ejected.

With these assumptions, the velocity of the shell once the gas is instantaneously released is given by
\begin{equation}
\label{velocity}
	v = - \frac{P_B^z}{M}\,.
\end{equation}
Of course, our simple minded model for pulsar kicks is far removed from a proper description of a fully developed model of core collapse supernova. Nevertheless, we hope that it captures the essential features of the propulsion mechanism to provide us with a reasonable estimate regarding its validity in describing pulsar kicks.

Typical masses for cores which collapse into pulsars are in the range of $1 \lesssim M/M_{\odot} \lesssim 2.5$, typical radii are $8 \lesssim r_*/\mbox{km} \lesssim 16$ and magnetic fields range from $10^{12}G$ for ordinary radio pulsars to over $10^{13}G$ in Magnetars. Theoretical estimates predict magnetic fields of order $10^{16}-10^{17}G$ in cores of young pulsars shortly after their birth. We refer the reader to \cite{Potekhin:2014hja} for a review. The electron chemical potential and lepton chemical potential are thought to be of the order of $200 MeV$ \cite{Prakash:1996xs}. Given that electron masses, however small, wash away any imbalance of electron chirality \cite{Grabowska:2014efa} implies that that $\mu^- \sim 70 MeV$.

\begin{figure}[hbt]
\begin{center}
	\includegraphics[scale=0.85]{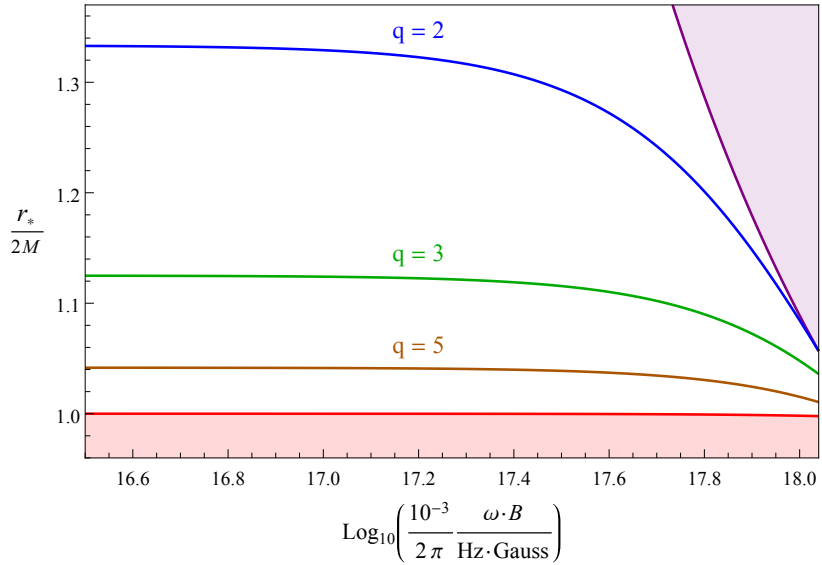}
	\protect\caption{\label{F:qvalues} Contours of equal $q =\frac{R^{2}+2r_{a}R+r_{s}^{2}}{R^{2}-r_{s}^{2}} $ towards the high end region of allowed values of magnetic fields. Once the radius drops below the Schwarzschild radius (red region) black holes are expected to form. Once the charge of the shell becomes larger than the mass (purple region), the line element is no longer real. In the allowed region, typical values of $q$ are of order unity.}
\end{center}
\end{figure}

To understand the dependence of \eqref{PBz} on the mass, $M$, radius, $r_*$, and magnetic field, $B$, of the rotating shell we need, in particular, to understand the ratio $ q=\frac{R^{2}+2r_{a}R+r_{s}^{2}}{R^{2}-r_{s}^{2}} $ where, we remind the reader, $r_a=M/2$, $r_s=\sqrt{M^2-Q^2}/2$, and $R$ and $Q$ are related to $r_*$ and $B$ via \eqref{Rstar}, \eqref{E:Bval} and \eqref{E:C1&C3}. Clearly, as the radius of the star gets closer to the Schwarzschild radius of the shell $q$ will grow without bound. This is similar to the ergosphere enhancement effect discussed in \cite{Shaverin:2014xya} and may be relevant for black hole kicks, see, e.g., \cite{Janka:2013hfa,Mandel:2015eta}. Contour plots of $q$ for relevant astrophysical data can be found in figure \ref{F:qvalues}. Since typical neutron stars have radii much larger than their Schwarzschild radius, we find that $q \sim 1$.

\begin{figure}[h]
	\centering\includegraphics[scale=1]{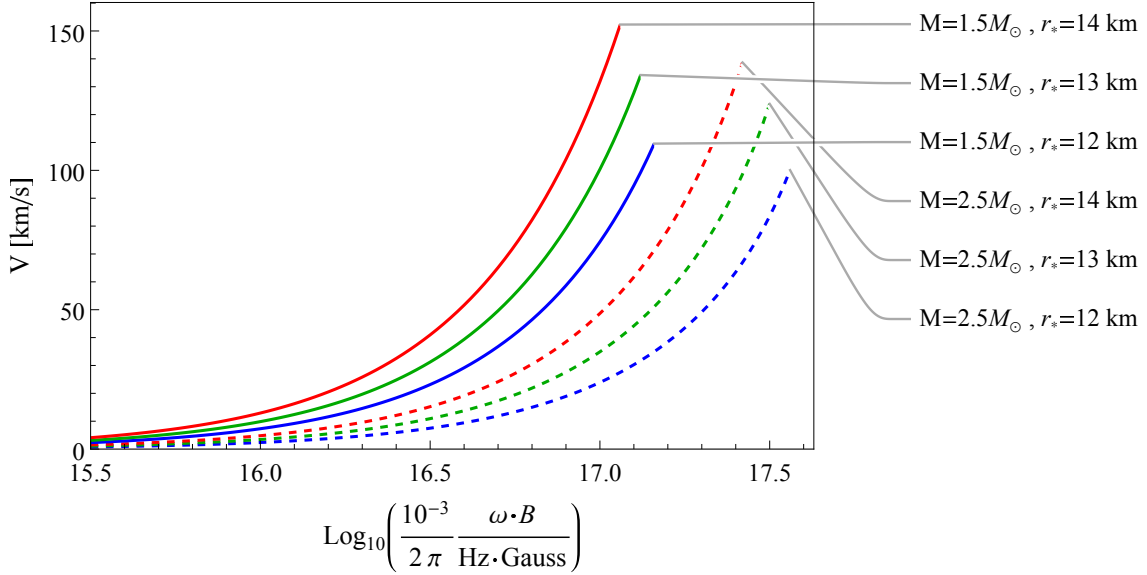}
	\protect\caption{\label{F:velocity} The dependence of the kick velocity on astrophysical parameters. }
\end{figure}

In figure \ref{F:velocity} we have plotted the kick velocity, given in equations \eqref{velocity} and \eqref{PBz} as a function of astrophysical data. 
To observe a kick at velocities of hundreds of kilometers per second magnetic fields and rotation rates of order of $10^{17}$ $\hbox{Gauss}\times\hbox{MegaHertz}$ are required. Such values of the magnetic field lie at the very edge of allowed magnetic fields in the interior of pulsars at their early stage of evolution, placing a question mark on the reliability of this mechanism in generating pulsar kicks.

We would like to point out that our estimate is rather different from that of \cite{Kaminski:2014jda} which allowed for a higher value of the kick. The difference between the results presented in figure \ref{F:velocity} and that of \cite{Kaminski:2014jda} stems from the difference in the propulsion mechanism. Here, we assumed that once the shell becomes transparent to neutrinos, i.e., the mean free path for neutrinos is larger than the radius of the shell, then those neutrinos which were in the core escape generating a balloon like mechanism which leads to values given in figure \ref{F:velocity}. Put differently, since new neutrinos can not thermalize during the escape process (due to their large mean free path) the total momentum available to them is that prior to the shell becoming transparent and so, the momentum gained by the shell is independent of the time scale during which the neutrinos are removed. The authors of \cite{Kaminski:2014jda} assumed a jet like mechanism where neutrinos are constantly being created as they are emitted. In this case, in order to estimate the momentum of the shell one can integrate the energy flux (equal to the momentum density) over a cross section of the core and over the time scale for emission which is roughly one second, $t_e \sim 1 \hbox{sec}$. Due to the enormously large ratio
\begin{equation}
	\frac{t_e}{r_*} \simeq 3\times10^4
\end{equation}
the estimates obtained by \cite{Kaminski:2014jda} differ from those in figure \ref{F:velocity}.

\section*{Acknowledgments}
The author is deeply grateful to his advisor, Amos Yarom, for his guidance, insightful suggestions, and his help in the revision of this paper. The author would also like to thank S. Bhattacharyya and H. Liu for useful discussions. This research was supported in part by the Israeli Science Foundation under an ISF excellence center grant 1989/14 and a BSF grant 2016324.

\begin{appendix}

\section{$U(1)$ anomalies for massless electrons.}
\label{A:triangle}
Consider the single family Lepton sector of the standard model with massless neutrinos.
\begin{equation}
\label{E:EW}
	L =i \bar{L}  \slashed{D} L + i \bar{e} \slashed{D} e - y_e h \bar{L} e + \ldots
\end{equation}
where $L = (e_L,\,\nu_L)$ is the left handed lepton multiplet, $e_R$ is the right handed electron, $h$ is the Higgs doublet and $\slashed{D} = \gamma^{\mu}D_{\mu}$ with $D_{\mu}$ the $SU(2)_w\times U(1)_Y$ covariant derivative. The dots denote additional terms which do not involve the Lepton sector. Note that \eqref{E:EW} has a global (Lepton number) $U(1)$ symmetry which rotates $L$ and $e_R$ by the same phase. If we set $y_e = 0$ then this $U(1)$ Lepton number symmetry is enhanced to $U(1)_\ell \times U(1)_r$ which we will refer to, following \cite{Kaminski:2014jda}, as left handed and right handed lepton number; $U(1)_r$ rotates $e_R$ by a phase and leaves $L$ invariant whereas $U(1)_\ell$ rotates $L$ by a phase and leaves $e_R$ invariant. 

The $U(1)_\ell \times U(1)_r$ symmetry of \eqref{E:EW} with $y_e=0$ is still realized linearly after electroweak symmetry breaking. In what follows we will refer to the appropriate classically conserved currents as $J_{\ell}^{\mu}$ and $J_{r}^{\mu}$ respectively. Our goal in this section is to compute the anomalous non-conservation laws for these currents. To this end, we need to consider the three point functions of the $U(1)$ currents in our theory.

Let us denote $I_{abc}^{\mu\nu\rho}(x,y,z) = \langle J_a^{\mu}(x)J_b^{\nu}(y)J_c^{\rho}(z)\rangle$ where $a,b,c = r,\,\ell,\,e$ referring to $U(1)_r$, $U(1)_\ell$ and the electroweak $U(1)$ respectively. The non-vanishing the three point functions which we wish to compute are 
$
	I_{r\,r\,r}^{\mu\nu\rho}\,,
	\,
	I_{e\,e\,r}^{\mu\nu\rho}\,,
	\,
	I_{r\,r\,e}^{\mu\nu\rho}\,,
	\,
	I_{\ell\,\ell\,\ell}^{\mu\nu\rho}\,,
	\,
	I_{e\,e\,\ell}^{\mu\nu\rho}
$
and
$
	I_{\ell\,\ell\,e}^{\mu\nu\rho}
$.

Following the notation of \cite{srednicki2007quantum}, let us denote
\begin{equation}
\label{E:ItoM}
	i \mathcal{M}_{abc}^{\mu\nu\rho} (2\pi)^4 \delta^{(4)}(p+q+k) = \int e^{i(px+qy+kz)} I_{abc}^{\mu\nu\rho}(x,y,z) d^4x d^4y d^4z\,.
\end{equation}
For a single chiral fermion with chiral current $J^{\mu}$ we have
\begin{align}
\begin{split}
\label{E:Mvals}
	p_{\mu}\mathcal{M}^{\mu\nu\rho} =& -\frac{f}{8\pi^2} (1-c)q_{\alpha}k_{\beta}\epsilon^{\nu\rho\alpha\beta} \\ 
	q_{\mu}\mathcal{M}^{\mu\nu\rho} =& -\frac{f}{8\pi^2} (1-c)k_{\alpha}p_{\beta}\epsilon^{\nu\rho\alpha\beta} \\ 
	k_{\mu}\mathcal{M}^{\mu\nu\rho} =& -\frac{f}{8\pi^2} 2 c \,p_{\alpha}q_{\beta}\epsilon^{\nu\rho\alpha\beta} \\ 
\end{split}
\end{align}
where $c$ is an ambiguity associated with shifting the momentum in the fermion loop which is linearly divergent, and $f=\sum_i \chi_i$ where $i$ runs over all fermion species in the loop, $\chi = 1$ for right handed fermions and $-1$ for left handed ones.

The amplitude $\mathcal{M}_{r\,r\,r}^{\mu\nu\rho}$ has a right handed fermion running in the loop so we should use $f=1$ and $c=1/3$ from symmetry. By requiring that the electromagnetic current is conserved one can obtain similar results which are summarized in table \ref{T:Ms} 
\begin{table}[hbt]
\begin{center}
\begin{tabular}{| c || c | c | c | c | c | c | }
\hline
	Amplitude & $\mathcal{M}_{r\,r\,r}$ & $\mathcal{M}_{e\,e\,r}$ &  $\mathcal{M}_{r\,r\,e}$  & $\mathcal{M}_{\ell\,\ell\,\ell}$ & $\mathcal{M}_{e\,e\,\ell}$ &$\mathcal{M}_{\ell\,\ell\,e}$  \\
\hline
	c & $\frac{1}{3}$ & $1$ & $0$ & $\frac{1}{3}$ & $1$ & $0$ \\
\hline
	f & $1$ & $1$ & $1$ & $-2$ & $-1$ & $-1$ \\
\hline
\end{tabular}
\caption{\label{T:Ms} Table of values used for $c$ and $f$ in \eqref{E:Mvals} for the various triangle diagrams computed in this section.}
\end{center}
\end{table}

To obtain the (non-)conservation law for the currents in the problem we couple $J_r^{\mu}$ and $J_{\ell}^{\mu}$ to external sources $A_{r}^{\mu}$ and $A_{\ell}^{\mu}$ with arbitrary couplings $q_{r}$ and $q_{\ell}$. Since the external source has no kinetic term, one can shift $q_{r}$ and $q_{\ell}$ by a field redefinition of $A_{r}^{\mu}$ or $A_{\ell}^{\mu}$. Expanding the action in the path integral in powers of $q_{e}$, $q_{r}$ and $q_{\ell}$ we have
\begin{equation}
	q_a \langle J_a^{\mu}(x) \rangle_{A\neq 0} 
		= \sum_{m=0} \frac{1}{m!} \left\langle \left(\sum_i q_i \int A_{i\,\nu} J_{i}^{\nu} d^4y \right)^{m}q_a J_a^{\mu}(x) \right\rangle_{A=0} \,.
\end{equation}

By construction the electromagnetic current is conserved,
\begin{equation}
	\partial_{\mu}J^{\mu}_e=0\,.
\end{equation}
For the right handed Lepton current we find that
\begin{align}
\begin{split}
	\partial_{\mu} J_r^{\mu}(x) = & q_e q_r \int A_{r\,\nu}(y) A_{e\,\rho}(z) \partial_{\mu}  I_{r\,r\,e}^{\mu\nu\rho}(x,y,z) d^4y d^4z + \frac{q_r^2}{2} \int  A_{r\,\nu}(y) A_{r\,\rho}(z) \partial_{\mu} I_{r\,r\,r}^{\mu\nu\rho}(x,y,z) d^4y d^4z 
		\\&+\frac{ q_e^2}{2} \int A_{e\,\nu}(y) A_{e\,\rho}(z) \partial_{\mu} I_{e\,e\,r}^{\nu\rho\mu}(y,z,x) d^4y d^4z
\end{split}
\end{align}

Inverting \eqref{E:ItoM} we find
\begin{align}
\begin{split}
\label{E:intI}
	\int A_{b\,\nu}(y) A_{c\,\rho}(z)  \partial_\mu I_{abc}^{\mu\nu\rho}(x,y,z) d^4y d^4z 
	&=  -\frac{1}{4} \frac{f}{8\pi^2} (1-c)  \epsilon^{\alpha\nu\beta\rho} F_{b\,\alpha\nu}(x) F_{c\,\beta\rho}(x)\,,\\
	\int A_{a\,\nu}(y) A_{b\,\rho}(z) \partial_{\mu} I_{a\,b\,c}^{\nu\rho\mu}(y,z,x) d^4y d^4z
	&=-\frac{1}{4} \frac{f}{8\pi^2} 2c\,  \epsilon^{\alpha\nu\beta\rho} F_{a\,\alpha\nu}(x) F_{b\,\beta\rho}(x)\,.
\end{split}
\end{align}
Thus,
\begin{multline}
		\partial_{\mu}J_r^{\mu} = - \frac{q_e q_r}{32\pi^2} \epsilon^{\alpha\nu\beta\rho} F_{e\,\alpha\nu}(x) F_{r\,\beta\rho}(x) - \frac{q_r^2}{3 \cdot 32\pi^2}  \epsilon^{\alpha\nu\beta\rho} F_{r\,\alpha\nu}(x) F_{r\,\beta\rho}(x)
		\\ - \frac{q_e^2}{32\pi^2}  \epsilon^{\alpha\nu\beta\rho} F_{e\,\alpha\nu}(x) F_{e\,\beta\rho}(x)\,.
\end{multline}
Similarly,
\begin{multline}
		\partial_{\mu}J_\ell^{\mu} = \frac{q_e q_\ell}{32\pi^2} \epsilon^{\alpha\nu\beta\rho} F_{e\,\alpha\nu}(x) F_{\ell\,\beta\rho}(x) + \frac{2 q_r^2}{3 \cdot 32\pi^2}  \epsilon^{\alpha\nu\beta\rho} F_{\ell\,\alpha\nu}(x) F_{\ell\,\beta\rho}(x)
		\\ + \frac{q_e^2}{32\pi^2}  \epsilon^{\alpha\nu\beta\rho} F_{e\,\alpha\nu}(x) F_{e\,\beta\rho}(x)\,.
\end{multline}

\section{Metric for a rotating charged shell.}
\label{A:rotating}

Consider a uniformly charged and uniformly rotating shell of mass $M$ (surface mass density $\sigma_M$), charge $Q$ (surface charge density $\sigma_Q$), angular velocity $\omega$ (as seen by an observer at infinity) and proper radius $r_*$. 
Our goal is to solve the Einstein equations
\begin{align}
	R_{\mu\nu} - \frac{1}{2}g_{\mu\nu}R &= \frac{8 \pi G}{c^4} \frac{1}{4\pi}\left(F_{\mu}{}^{\alpha}F_{\alpha}{}^{\nu} - \frac{1}{4} g^{\mu\nu}F_{\alpha\beta}F^{\alpha\beta}\right) + \frac{8 \pi G}{c^4} T^{\mu\nu}_{shell}\\
	\nabla_{\mu}F^{\mu\nu} & = \frac{4\pi}{c}J^{\nu}_{shell}
\end{align}
where $T^{\mu\nu}_{shell} = \tau^{\mu\nu}\delta(r-R)$ and $J^{\mu}{}_{shell}=j^{\mu} \delta(r-R)$ are the stress tensor and charge current of the rotating shell. As discussed in \cite{israel1967nuovo}  continuity of the Einstein and Maxwell equations then demand that
\begin{align}
\label{E:IJC}
\begin{split}
	 -\left[K^{\mu\nu} - K h^{\mu\nu} \right] & = \frac{8\pi G}{c^4}\tau^{\mu\nu} \\
	 \left[n_{\alpha}F^{\alpha\mu} \right] & = \frac{4\pi}{c}j^{\mu} 
\end{split}
\end{align}
where $h^{\mu\nu}=g^{\mu\nu} -n^{\mu}n^{\nu}$ is the induced metric on the shell with $n^{\mu}$ an outward pointing normal vector to the shell, $K^{\mu}{}_{\nu} = h^{\mu\gamma}\nabla_{\gamma}n_{\nu}$ is the induced metric on the shell whose indices are raised and lowered with the metric $g_{\mu\nu}$ and $K = K^{\mu}{}_{\mu}$. Square brackets denote the difference between quantities slightly outside the shell and slightly inside the shell. In what follows we will set $G=c=1$. 

Since the shell is uniformly rotating, its 4-velocity $u^{\mu}$ is given by $u^{\phi} = \omega u^t$ with all other components vanishing. We then define the energy density $\sigma_M$ and charge density $\sigma_Q$ via 
\begin{equation}
\label{E:sigmaeqs}
	\tau^{\mu\nu}u_{\nu} = -\sigma_M u^{\mu} \qquad j^{\mu}=\sigma_Q u^{\mu}\,.
\end{equation}
We can expand $\sigma_M$ and $\sigma_Q$ in powers of $\omega$ and solve \eqref{E:sigmaeqs} order by order in $\omega$.
Our ansatz for the metric and gauge field is given by
\begin{align}
\begin{split}
	ds^2 &= -e^{2 U}dt^2 + e^{-2 U}\left(e^{2 K}(dr^2 + r^2 d\theta^2) + W^2 (d\phi - \omega A dt)^2 \right) \\
	F &= E_r dt \wedge dr + \omega r E_\theta dt \wedge d\theta +\omega r  \sin\theta  B_{\theta} d\phi \wedge dr + \omega r^2 \sin\theta B_r d\phi \wedge d\theta\,,
\end{split}
\end{align}
which are also expanded perturbatively in $\omega$,
\begin{align}
\begin{split}
\label{E:wexpansion}
	U &= U^{(0)} + \mathcal{O}(\omega^2) 
	\quad
	K = K^{(0)} + \mathcal{O}(\omega^2)
	\quad
	W = W^{(0)} + \mathcal{O}(\omega^2)
	\\
	A &= A^{(0)} + \mathcal{O}(\omega^2)
	\quad
	E_i = E_i^{(0)} + \mathcal{O}(\omega^2)
	\quad
	B_i = B_i^{(0)} + \mathcal{O}(\omega^2) \,.
\end{split}
\end{align}

To leading order in $\omega$ we find that the solution in the exterior of the shell is that of a charged black hole and in the interior of a rotating flat space-time,
\begin{align}
\label{E:0ordersol}
\begin{split}
U^{(0)} &= \begin{cases} 
\ln \left(\frac{r^2-r_s^2}{r^2 + 2 r_a r +r_s^2}\right) & r \geq R \\
\ln \left(\frac{R^2-r_s^2}{R^2 + 2 r_a R +r_s^2}\right) & r < R
\end{cases}
\qquad
K^{(0)} = \begin{cases} 
\ln \left(\frac{r^2-r_s^2}{r^2}\right) & r \geq R \\
\ln \left(\frac{R^2-r_s^2}{R^2}\right) & r < R
\end{cases}
\\
E_r &= \begin{cases} \frac{2c^{2}\sqrt{r_{a}^{2}-r_{s}^{2}}\left(r^{2}-r_{s}^{2}\right)}{\sqrt{G_{N}}\left(2rr_{a}+r^{2}+r_{s}^{2}\right)^{2}} & r \geq R \\
0 & r<R 
\end{cases}
\end{split}
\end{align}

together with
\begin{equation}
	W = e^{K^{(0)}}r\sin\theta\,.
\end{equation}
Here, the parametric radius of the shell is related to its proper radius via $r_* = \int_0^R \sqrt{g_{rr}}dr$.  
The parameters $r_s$ and $r_a$ are integration constants which specify the mass density and charge density, 
\begin{equation}
	\sigma_M^{(0)} = \frac{c^{4}R\left(r_{a}R+r_{s}^{2}\right)}{2\pi G_{N}\left(2r_{a}R+r_{s}^{2}+R^{2}\right)^{2}}
	\qquad
	\sigma_Q^{(0)}=\frac{c^{2}R^{2}\sqrt{r_{a}^{2}-r_{s}^{2}}}{2\pi\sqrt{G_{N}}\left(2r_{a}R+r_{s}^{2}+R^{2}\right)^{2}}\,.
\end{equation}
By comparing \eqref{E:0ordersol} to the Reissner-Nordstr\"om black hole solution we find that 
\begin{equation}
	r_a = M/2
	\qquad
	r_s = \frac{1}{2} \sqrt{M^2 - Q^2}\,.
\end{equation}

At first order in $\omega$ we solve the equations separately in the interior and exterior and then glue them using Israel's junction condition. We start with the interior. 
In the interior, the equation for $E^{(0)}_\theta$ is given by
\begin{equation}
	\partial_\theta (E^{(0)}_\theta \sin\theta) = 0
	\qquad
	\partial_r (E^{(0)}_\theta r) = 0
\end{equation}
whose solution is
\begin{equation}
	E^{(0)}_\theta \propto \frac{1}{r}
\end{equation}
which is divergent. Thus, we set $E^{(0)}_\theta=0$. 
The equation for $B_i^{(0)}$ is given by
\begin{align}
\begin{split}
	\sin\theta \partial_r (r^2 B_r^{(0)}) + r \partial_{\theta}(\sin\theta B^{(0)}_{\theta}) &= 0 \\
	\partial_r (r B^{(0)}_{\theta})-\partial_{\theta}(B^{(0)}_r)  & = 0\,.
\end{split}
\end{align}
Using separation of variables and requiring that the solution is not divergent near $r=0$, $\theta=0$ and $\theta=\pi$, we find that
\begin{equation}
\label{E:BsolV1}
	B_r^{(0)} = C_1 \cos\theta + \frac{C_2}{6} r (1+3 \cos(2\theta)) 
	\qquad
	B_{\theta}^{(0)} = -C_1 \sin\theta - C_2 r \cos\theta \sin\theta \,.
\end{equation}
with $C_1$ and $C_2$ integration constants.
Solving for $A^{(0)}$ one finds, using separation of variables, an entire family of solutions,
\begin{equation} \label{E:Ainterior}
	A^{(0)} = C_3 + C_4 r \cos\theta + C_5 r^2 (1-5 \cos^2\theta) + \mathcal{O}(r^3)\,.
\end{equation}
The undetermined coefficients, $C_i$ can be computed by matching the interior and exterior solution across the shell. 

Before proceeding with the exterior solution which is somewhat more involved, it is useful to study the implications of the Israel junction condition to simplify the interior solution \eqref{E:Ainterior} and \eqref{E:BsolV1}. Let us denote $A_i(r,\theta) = A^{(0)}(r<R,\theta)$ and $A_e(r,\theta) = A^{(0)}(r>R,\theta)$, then the matching condition for the metric \eqref{E:IJC} take the form 
\begin{equation} \label{E:IJmetricO1}
	\partial_r \left(A_i(r,\theta) - A_e(r,\theta)\right)\big|_{r=R} + Q_1 A_i(R,\theta)  = Q_2
\end{equation}
where the $Q_i$ are constant ($\theta$ and $r$ independent) coefficients whose explicit form is somewhat long and irrelevant to our construction. On physical grounds we expect that there is no non-trivial solution in the absence of a shell. Thus, \eqref{E:Ainterior} gets truncated to
\begin{equation}
	A^{(0)} = C_3
\end{equation} 
and the exterior solution for $A$ must be $\theta$ independent. 
Likewise,
if we denote by $B_i(r,\theta) = B_\theta^{(0)}(r<R,\theta)$ and $B_e(r,\theta) = B_\theta^{(0)}(r>R,\theta)$ then the matching conditions for the field strength \eqref{E:IJC} are of the form
\begin{equation} \label{E:IJfieldO1}
	B_e(R,\theta) - B_i(R,\theta) + Q_3 A_i(R,\theta) \sin\theta= Q_4 \sin\theta
\end{equation}
where $Q_3$ and $Q_4$ are constant (and we have consistently taken $E_\theta=0$ in the exterior). By the same arguments used to study $A^{(0)}$ we find that $C_2$ in \eqref{E:BsolV1} must vanish. Thus, in the interior we have:
\begin{equation} \label{E:interiorO1}
	A^{(0)} = C_3
	\qquad
	B_r^{(0)} = C_1 \cos\theta 
	\qquad
	B_{\theta}^{(0)} = -C_1 \sin\theta
\end{equation}
with vanishing correction to $E_{\theta}$.

Turning to the exterior, the local (in $\theta$) nature of \eqref{E:IJmetricO1} and \eqref{E:IJfieldO1} imply that in the exterior the solution must take the form
\begin{equation}
	A^{(0)} = \alpha(r)
	\qquad
	B_r^{(0)} = \frac{c^{2}}{\sqrt{G_{N}}} \beta(r) \cos\theta
	\qquad
	B_{\theta}^{(0)} = -\frac{c^{2}}{\sqrt{G_{N}}}\frac{1}{2r} \partial_r \left(r^2 \beta(r)\right) \sin\theta
\end{equation}
where the last term is determined via the Bianchi identity. The equations for $\alpha$ and $\beta$ form a coupled set of second order linear equations of the form,
\begin{subequations}
\label{E:abeq}
\begin{equation}
	\left( \frac{(r^2+2 r r_a+r_s^2)^4}{r^2 (r^2-r_s^2)} \alpha' - 4 r^2 \sqrt{r_a^2 - r_s^2} \beta \right)' = 0 
\end{equation}
\begin{multline}
	\beta'' + \left(\frac{6}{r} + \frac{2r}{r^2-r_s^2} - \frac{4(r+r_a)}{r^2+2 r r_a+r_s^2} \right) \beta'
	+\frac{4(2r^3 r_a+3 r^2 r_s^2 -r_s^4)}{r^2(r^2-r_s^2)(r^2+2 r r_a +r_s^2)}\beta \\
	- \frac{4 \sqrt{r_a^2-r_s^2}(r^2+2 r r_a+r_s^2)^2}{r^4(r^2-r_s^2)}\alpha' =0
\end{multline}
\end{subequations}

Fortunately \eqref{E:abeq} can be solved analytically, though the explicit form of the solution is somewhat long. The most general solution to \eqref{E:abeq} contains four integration constant two of which are determined by requiring that the metric be asymptotically flat. The other two integration constants, as well as $C_3$ and $C_1$ in \eqref{E:interiorO1} are obtained by solving \eqref{E:IJC} together with continuity of the metric and field strength in the transverse directions. We find
\begin{align}
\begin{split}
\label{E:C1&C3}
C_1 &= -\frac{c^{2}}{\sqrt{G_{N}}}\frac{\sqrt{\rho_{a}^{2}-1}\left(1+2\rho_{a}\rho_{R}+\rho_{R}^{2}\right)^{2}\left(2\rho_{R}P_{a}+3\rho_{a}\left(\rho_{R}^{2}-1\right)^{3}P_{b}\log\left(\frac{\rho_{R}-1}{\rho_{R}+1}\right)\right)}{\rho_{R}^{4}\left(2\rho_{R}Q_{a}+\left(\rho_{R}^{2}-1\right)Q_{b}\log\left(\frac{\rho_{R}-1}{\rho_{R}+1}\right)\right)}  \\
C_3 &= \frac{2\rho_{R}S_a+\left(\rho_{R}^{2}-1\right)S_b\log\left(\frac{\rho_{R}-1}{\rho_{R}+1}\right)}{\left(1+2\rho_{a}\rho_{R}+\rho_{R}^{2}\right)\left(2\rho_{R}Q_{a}+\left(\rho_{R}^{2}-1\right)Q_{b}\log\left(\frac{\rho_{R}-1}{\rho_{R}+1}\right)\right)}
\end{split}
\end{align}
where
\begin{align}
\begin{split}
P_a &= \rho_{a}^{2}\left(3\rho_{R}^{6}+\rho_{R}^{4}+21\rho_{R}^{2}-9\right)\rho_{R}+\rho_{a}\left(9\rho_{R}^{6}+27\rho_{R}^{4}-\rho_{R}^{2}-3\right)+16\rho_{R}^{5} \\
P_b &= \rho_{a}\left(\rho_{R}^{2}+3\right)\rho_{R}+3\rho_{R}^{2}+1 \\
Q_a & = 12\rho_{a}^{2}\left(\rho_{R}^{2}-3\right)\rho_{R}^{4}+\rho_{a}^{3}\left(3\rho_{R}^{6}+\rho_{R}^{4}-15\rho_{R}^{2}+3\right)\rho_{R}-12\rho_{a}\left(\rho_{R}^{5}+\rho_{R}^{3}\right)-7\rho_{R}^{6}+3\rho_{R}^{4}-5\rho_{R}^{2}+1 \\
Q_b & = 12\rho_{a}^{2}\left(\rho_{R}^{2}+3\right)\rho_{R}^{4}+\rho_{a}^{3}\left(3\rho_{R}^{6}+3\rho_{R}^{4}+13\rho_{R}^{2}-3\right)\rho_{R}+12\rho_{a}\left(3\rho_{R}^{5}+\rho_{R}^{3}\right)+\left(\rho_{R}^{2}+1\right){}^{2}\left(5\rho_{R}^{2}-1\right) \\
S_a  & =2\rho_{a}^{4}\left(3\rho_{R}^{6}+\rho_{R}^{4}-15\rho_{R}^{2}+3\right)\rho_{R}^{2}+\rho_{a}\left(\rho_{R}^{2}+1\right){}^{2}\left(3\rho_{R}^{4}-24\rho_{R}^{2}+5\right)\rho_{R}\\&\phantom{=}+\rho_{a}^{2}\left(-3\rho_{R}^{8}+10\rho_{R}^{6}-128\rho_{R}^{4}+22\rho_{R}^{2}+3\right)+\rho_{a}^{3}\left(-3\rho_{R}^{9}+38\rho_{R}^{7}-76\rho_{R}^{5}-38\rho_{R}^{3}+15\rho_{R}\right)\\&\phantom{=}+5\rho_{R}^{8}-32\rho_{R}^{6}+18\rho_{R}^{4}-8\rho_{R}^{2}+1 \\
S_b & =\rho_{a}^{4}\left(6\rho_{R}^{8}+6\rho_{R}^{6}+26\rho_{R}^{4}-6\rho_{R}^{2}\right)+\rho_{a}^{3}\left(-3\rho_{R}^{9}+36\rho_{R}^{7}+82\rho_{R}^{5}+28\rho_{R}^{3}-15\rho_{R}\right)\\&\phantom{=}-3\rho_{a}^{2}\left(\rho_{R}^{8}-40\rho_{R}^{6}-34\rho_{R}^{4}+8\rho_{R}^{2}+1\right)+\rho_{a}\rho_{R}\left(3\rho_{R}^{8}+16\rho_{R}^{6}+114\rho_{R}^{4}-5\right)\\&\phantom{=}+\left(\rho_{R}^{2}+1\right){}^{3}\left(5\rho_{R}^{2}-1\right) \,,
\end{split}
\end{align}
and we have defined
\begin{equation}
	\rho_a = \frac{r_a}{r_s}
	\qquad
	\rho_R = \frac{R}{r_s} \,.
\end{equation}

\section{Anomalies and transport}
\label{A:polynomial}
In what follows we use the general prescription of \cite{Jensen:2013rga} to compute the contribution of anomalies to transport. Our starting point is the anomaly polynomial
\begin{multline} \label{E:AnomalyPolynomial}
	\mathcal{P} = \frac{c_{1}}{3}F_{r}\wedge F_{r}\wedge F_{r}+c_{2}F_{e}\wedge F_{r}\wedge F_{r}+c_{3}F_{r}\wedge F_{e}\wedge F_{e}+c_{m}^{1}F_{r}\wedge R\wedge R 
	\\
	+ \frac{c_{4}}{3}F_{\ell}\wedge F_{\ell}\wedge F_{\ell}+c_{5}F_{e}\wedge F_{\ell}\wedge F_{\ell}+c_{6}F_{\ell}\wedge F_{e}\wedge F_{e}+c_{m}^{2}F_{\ell}\wedge R\wedge R 
	\,.
\end{multline}
with
\begin{equation} 
\label{ci_parameters}
\begin{matrix}
c_{1}=-c q_{r}^{3} & \qquad \,& c_{4}=2 c q_{l}^{3}\\
c_{2}=-cq_{e} q_{r}^{2} & \qquad & c_{5}=cq_{e} q_{l}^{2}\\
c_{3}=-cq_{e}^{2} q_{r} & \qquad & c_{6}=cq_{e}^{2} q_{l}\,,
\end{matrix}
\end{equation}
and $ c=\frac{1}{8\pi^{2}} $, responsible for the left and right handed lepton anomaly computed in appendix \ref{A:triangle}. The parameters $c_{m_i}$ are associated with mixed gauge-gravitational anomalies. They do not play a role in the analysis presented in this work. Their value can be computed by standard means.

To apply the prescription of \cite{Jensen:2013rga} we construct the thermal anomaly polynomial
\begin{equation} \label{E:RepRul} \nonumber
\mathcal{P}_{T}=\mathcal{P}\left(F_r,F_\ell,F_e,\,p_{k}\left(R\right)\rightarrow p_{k}\left(R\right)-\frac{F_{T} \wedge F_{T}}{4\pi^2}\wedge p_{k-1}\left(R\right)\right)
\end{equation}
with $F_T$ an auxiliary field strength which will be set to zero at the end of the computation and $p_k$ the $k$th Pontryagin class. For the particular case at hand we need only
$p_{1}\left(R\right)=-\frac{1}{8\pi^{2}}R\wedge R $. 
From the thermal anomaly polynomial we construct the hatted thermal anomaly polynomial,
\begin{equation}
	\hat{\mathcal{P}}_T(F_{\ell},\,F_{r},\,F_{e},\,F_T,\, R) = \mathcal{P}_T \left(\hat{F}_{\ell},\,\hat{F}_{r},\,\hat{F}_{e},\,\hat{F}_T,\, \hat{R} \right)\,,
\end{equation}
where hatted two-forms $\hat{F}=d\hat{A}$ and $R = d\hat{\Gamma}+ \hat{\Gamma}^2$ are related to their unhated counterpart $\hat{F}=d\hat{A}$ and $R = d\hat{\Gamma}+ \hat{\Gamma}^2$ via a shift,
\begin{equation}
	\hat{A} = A + \mu u
	\qquad
	\hat{\Gamma} = \Gamma + \mu_B u
\end{equation}
with $\mu$ the chemical potential associated with the relevant flavor field (and $\mu_T = 2\pi T$ with $T$ the temperature), $u$ the velocity 1-form $u=u_{\mu}dx^{\mu}$ with $u^{\mu}$ the velocity field, and $(\mu_R)^{\alpha}{}_{\beta} = T \nabla_{\alpha} \left(u^{\alpha}/T\right)$ the spin chemical potential. From $\mathcal{P}_T$ and $\hat{\mathcal{P}}_T$ we construct the master function
\begin{equation} \label{E:VT}
V_{T}=\frac{u}{2\omega}\wedge\left(\mathcal{P}_{T}-\hat{\mathcal{P}}_{T}\right) \,,
\end{equation}
from which the conserved currents can be computed via a variational prescription. (Note that the $1/\omega$ in \eqref{E:VT} is a shorthand notation for removing one factor of $\omega$ from $u \wedge \left(\mathcal{P}_{T}-\hat{\mathcal{P}}_{T}\right) $ which can be shown to vanish when $\omega$ is set to zero.) More precisely, we have
\begin{equation}
\label{E:currentfromV}
	\star J = \frac{\partial V_T}{\partial \mathcal{B}}
	\qquad
	\star q =\frac{1}{2} \frac{\partial V_T}{\partial \omega}
	\qquad
	\star L^{\mu}{}_{\nu} =\frac{\partial V_{T}}{\partial\left(\mathcal{B}_{R}\right)^{\nu}_{\mu}} \,.
\end{equation}
Here $J = J_{cov\,\mu}dx^{\mu}$ with $J_{cov}^{\mu}$ the covariant current and $\mathcal{B} = P_{\mu}{}^{\alpha}P_{\nu}{}^{\beta}F_{\alpha \beta}dx^{\mu}dx^{\nu}$ is the magnetic field associated with the field strength $F$ with $P^{\mu\nu} = g^{\mu\nu} + u^{\mu}u^{\nu}$, $q = q_{\mu}dx^{\mu}$ and $L^{\mu}{}_{\nu} = L^{\mu}{}_{\nu\alpha} dx^{\alpha}$ specify the anomalous contribution to the covariant stress tensor
\begin{equation}
	T^{\mu\nu}_{cov} = u^{\mu}q^{\nu} + u^{\nu}q^{\mu} + \nabla_{\rho} \left( L^{\mu[\nu\rho]} + L^{\nu[\mu\rho]} - L^{\rho(\mu\nu)}\right)
\end{equation}
with square brackets denoting antisymmetrization of the indices and circular brackets their symmetrization,
\begin{equation} 
	A^{\left(\mu\nu\right)}=\frac{1}{2}\left(A^{\mu\nu}+A^{\nu\mu}\right)\,,\quad A^{\left[\mu\nu\right]}=\frac{1}{2}\left(A^{\mu\nu}-A^{\nu\mu}\right) \,.
\end{equation}
The magnetic component of the Riemann tensor two form $(\mathcal{B}_R)^{\mu}{}_{\nu}$ is given by
\begin{equation}
	(B_R)^{\mu}{}_{\nu} = R^{\mu}{}_{\nu\alpha\beta} P^{\alpha}{}_{\rho}P^{\beta}{}_{\sigma}dx^{\rho}dx^{\sigma}\,.
\end{equation}

Using the above prescription we find, after some algebra,
\begin{align}
\begin{split}
	J_{r,cov}^{\mu} = & -2\left(c_{2}\mu+3c_{1}\mu_{r}\right)B_{r}^{\mu}-2\left(c_{3}\mu+c_{2}\mu_{r}\right)B^{\mu}-2c_{m}^{1}\left(\mu_{R}\right)_{\phantom{\alpha}\beta}^{\alpha}\left(B_{R}\right)_{\phantom{\mu\beta}\alpha}^{\mu\beta} \\
 	&  -\left(8\pi^{2}T^{2}c_{m}^{1}+c_{3}\mu^{2}+2c_{2}\mu\mu_{r}+3c_{1}\mu_{r}^{2}+c_{m}^{1}\mbox{tr}\left(\mu_{R}^{2}\right)\right)\omega^{\mu} \,, \\
	J_{\ell,cov}^{\mu} = & -2\left(c_{5}\mu+3c_{4}\mu_{\ell}\right)B_{\ell}^{\mu}-2\left(c_{6}\mu+c_{5}\mu_{\ell}\right)B^{\mu}-2c_{m}^{2}\left(\mu_{R}\right)_{\phantom{\alpha}\beta}^{\alpha}\left(B_{R}\right)_{\phantom{\mu\beta}\alpha}^{\mu\beta}  \\
	& -\left(8\pi^{2}T^{2}c_{m}^{2}+c_{6}\mu^{2}+2c_{5}\mu\mu_{\ell}+3c_{4}\mu_{\ell}^{2}+c_{m}^{2}\mbox{tr}\left(\mu_{R}^{2}\right)\right)\omega^{\mu} \,, \\
	J_{e,cov}^{\mu} = & -2\left(c_{6}\mu+c_{5}\mu_{\ell}\right)B_{\ell}^{\mu}-2\left(c_{3}\mu+c_{2}\mu_{r}\right)B_{r}^{\mu}-2\left(c_{6}\mu_{\ell}+c_{3}\mu_{r}\right)B^{\mu} \\
	& -\left(2c_{6}\mu\mu_{\ell}+c_{5}\mu_{\ell}^{2}+2c_{3}\mu\mu_{r}+c_{2}\mu_{r}^{2}\right)\omega^{\mu} \,.
\end{split}
\end{align}
Here the magnetic fields $ B^{\mu} $ and $ \left(B_{R}\right)_{\phantom{\mu\alpha}\beta}^{\mu\alpha} $ are given by
\begin{equation} \nonumber
	B^{\mu}=\frac{1}{2}\epsilon^{\mu\nu\rho\sigma}u_{\nu}F_{\rho\sigma}\,,\qquad\left(B_{R}\right)_{\phantom{\mu\alpha}\beta}^{\mu\alpha}=\frac{1}{2}\epsilon^{\mu\nu\rho\sigma}u_{\nu}R_{\phantom{\alpha}\beta\rho\sigma}^{\alpha}\,,
\end{equation}
and should be distinguished from the magnetic field two forms $\mathcal{B}$ and $\mathcal{B}_R$ in \eqref{E:currentfromV}. 

The consistent currents are obtained from the covariant ones via $ J_{i,anom}^{\mu} = J_{i,cov}^{\mu} - J_{i,BZ}^{\mu} $, where $ J_{i,BZ}^{\mu} $ are Bardeen-Zumino terms given by $ J_{i,BZ} = \frac{\partial I_{CS}}{\partial F_i} $ where $I_{CS}$ is the Chern Simons term associated with the anomaly polynomial \eqref{E:AnomalyPolynomial},
\begin{align} \label{E:CSterm}
	I_{CS}= & \, \frac{c_{1}}{3}A_{r}\wedge F_{r}\wedge F_{r}+c_{2}A_{r}\wedge F_{e}\wedge F_{r}+c_{3}A_{r}\wedge F_{e}\wedge F_{e}+\left(c_{m}^{1}-s_{m}^{1}\right)A_{r}\wedge R\wedge R+s_{m}^{1}F_{r}\wedge i_{CS} \nonumber \\
	& +\frac{c_{4}}{3}A_{\ell}\wedge F_{\ell}\wedge F_{\ell}+c_{5}A_{\ell}\wedge F_{e}\wedge F_{\ell}+c_{6}A_{\ell}\wedge F_{e}\wedge F_{e}+\left(c_{m}^{2}-s_{m}^{2}\right)A_{\ell}\wedge R\wedge R+s_{m}^{2}F_{\ell}\wedge i_{CS} \,, \nonumber \\
\end{align}
with $ i_{CS}=\Gamma\wedge d\Gamma+\frac{2}{3}\Gamma\wedge\Gamma\wedge\Gamma $.
An explicit computation yields
\begin{align}
\begin{split}
	J_{r,BZ}^{\alpha}=&\frac{1}{2}\epsilon^{\alpha\beta\gamma\delta}\left(2c_{1}A_{r\beta}F_{r,\gamma\delta}+c_{2}A_{r\beta}F_{e,\gamma\delta}+s_{m}^{1}i_{CS}^{\alpha}\right)  \\
	J_{\ell,BZ}^{\alpha}=&\frac{1}{2}\epsilon^{\alpha\beta\gamma\delta}\left(2c_{4}A_{\ell\beta}F_{\ell,\gamma\delta}+c_{5}A_{\ell\beta}F_{e,\gamma\delta}+s_{m}^{2}i_{CS}^{\alpha}\right)  \\
	J_{e,BZ}^{\alpha}=&\frac{1}{2}\epsilon^{\alpha\beta\gamma\delta}\left(c_{2}A_{r\beta}F_{r,\gamma\delta}+2c_{3}A_{r\beta}F_{e,\gamma\delta}+c_{5}A_{\ell\beta}F_{\ell,\gamma\delta}+2c_{6}A_{\ell\beta}F_{e,\gamma\delta}\right) \,.
\end{split}
\end{align}

To find the constitutive relation of (the contribution of the anomaly to) the covariant stress-energy tensor, we evaluate $ q  $ and $L^{\mu}{}_{\nu}$ using $ \eqref{E:VT} $. We find
\begin{align}
\begin{split}
\label{E:qMU}
	q^{\mu}= & -2\left(c_{m}^{1}\mu_{r}\mbox{tr}\left(\mu_{R}^{2}\right)+c_{m}^{2}\mu_{\ell}\mbox{tr}\left(\mu_{R}^{2}\right)+c_{1}\mu_{r}^{3}+c_{2}\mu\mu_{r}^{2}+c_{3}\mu^{2}\mu_{r} \right. \nonumber \\ 	& \left. +c_{4}\mu_{\ell}^{3}+c_{5}\mu\mu_{\ell}^{2}+c_{6}\mu^{2}\mu_{\ell}+8\pi^{2}c_{m}^{1}\mu_{r}T^{2}+8\pi^{2}c_{m}^{2}\mu_{\ell}T^{2}\right)\omega^{\mu}  \\
	& -\left(c_{m}^{1}\mbox{tr}\left(\mu_{R}^{2}\right)+3c_{1}\mu_{r}^{2}+2c_{2}\mu\mu_{r}+c_{3}\mu^{2}+c_{m}^{1}8\pi^{2}T^{2}\right)B_{r}^{\mu} \nonumber \\ & -\left(c_{m}^{2}\mbox{tr}\left(\mu_{R}^{2}\right)+3c_{4}\mu_{\ell}^{2}+2c_{5}\mu\mu_{\ell}+c_{6}\mu^{2}+c_{m}^{2}8\pi^{2}T^{2}\right)B_{\ell}^{\mu}  \\
	& -\left(c_{2}\mu_{r}^{2}+2c_{3}\mu\mu_{r}+c_{5}\mu_{\ell}^{2}+2c_{6}\mu\mu_{\ell}\right)B^{\mu}-2\left(c_{m}^{1}\mu_{r}+c_{m}^{2}\mu_{\ell}\right)\left(\mu_{R}\right)_{\phantom{\alpha}\beta}^{\alpha}\left(B_{R}\right)_{\phantom{\mu\beta}\alpha}^{\mu\beta}\,.  \\
\end{split}
\end{align}
and
\begin{align}
	L_{\phantom{\mu\beta}\alpha}^{\mu\beta}=-2c_{m}^{1}B_{r}^{\mu}\left(\mu_{R}\right)_{\phantom{\beta}\alpha}^{\beta}-2c_{m}^{2}B_{\ell}^{\mu}\left(\mu_{R}\right)_{\phantom{\beta}\alpha}^{\beta}-2\left(c_{m}^{1}\mu_{r}+c_{m}^{2}\mu_{\ell}\right)\left(B_{R}\right)_{\phantom{\mu\beta}\alpha}^{\mu\beta}-2\left(c_{m}^{1}\mu_{r}+c_{m}^{2}\mu_{\ell}\right)\omega^{\mu}\left(\mu_{R}\right)_{\phantom{\beta}\alpha}^{\beta} \, .
\end{align}

Much like the consistent currents, the covariant stress tensor can be obtained from the covariant one by adding to it a Bardeen Zumino polynomial.
One finds
\begin{equation} \label{T_anom_UU}
	T^{\mu\nu}_{anom}=u^{\mu}q^{\nu}+u^{\nu}q^{\mu}+\nabla_{\lambda}\left(L^{\mu\left[\nu\lambda\right]}+L^{\nu\left[\mu\lambda\right]}-L^{\lambda\left(\mu\nu\right)}\right)+\frac{1}{2}\nabla_{\lambda}\left(X^{\lambda\mu\nu}+X^{\lambda\nu\mu}-X^{\mu\nu\lambda}\right) ,
\end{equation}
with $ X_{\phantom{\mu}\nu}^{\mu}=\frac{\partial I_{CS}}{\partial R_{\phantom{\nu}\mu}^{\nu}} $.
An explicit computation yields
\begin{align}
	X_{\phantom{\mu\lambda}\nu}^{\mu\lambda}  = & \frac{s_{m}^{1}}{2}\left[\epsilon^{\mu\rho\kappa\sigma}\Gamma_{\phantom{\lambda}\nu\rho}^{\lambda}+\epsilon^{\lambda\rho\kappa\sigma}\Gamma_{\phantom{\mu}\nu\rho}^{\mu}\right]F_{r,\kappa\sigma}+\frac{s_{m}^{2}}{2}\left[\epsilon^{\mu\rho\kappa\sigma}\Gamma_{\phantom{\lambda}\nu\rho}^{\lambda}+\epsilon^{\lambda\rho\kappa\sigma}\Gamma_{\phantom{\mu}\nu\rho}^{\mu}\right]F_{\ell,\kappa\sigma} \nonumber \\
	& +\left(c_{m}^{1}-s_{m}^{1}\right)\left[\epsilon^{\mu\rho\kappa\sigma}R_{\phantom{\lambda}\nu\kappa\sigma}^{\lambda}+\epsilon^{\lambda\rho\kappa\sigma}R_{\phantom{\mu}\nu\kappa\sigma}^{\mu}\right]A_{r,\rho}+\left(c_{m}^{2}-s_{m}^{2}\right)\left[\epsilon^{\mu\rho\kappa\sigma}R_{\phantom{\lambda}\nu\kappa\sigma}^{\lambda}+\epsilon^{\lambda\rho\kappa\sigma}R_{\phantom{\mu}\nu\kappa\sigma}^{\mu}\right]A_{\ell,\rho} \,. \nonumber \\
\end{align}
The resulting anomalous contribution to the stress tensor, to first order in derivatives (to which $L^{\mu\nu}_{\alpha}$ and $X^{\mu\nu}_{\alpha}$ don't contribute) can is given in \eqref{E:Tmnanomaly} and \eqref{E:q_first_order}.

\end{appendix}

\bibliographystyle{JHEP}
\bibliography{KAnom2}

\end{document}